\documentclass[dvips]{article}

\usepackage{icrc2011}
\usepackage{amsmath}
\usepackage{amssymb}
\usepackage{graphicx}

\title {Semi-Analytic Model for Intergalactic Gamma-Ray Cascades in Extragalactic Magnetic Fields}

\newcommand{\etal}{\MakeLowercase{\textit{et al. }}} 
\shorttitle{H.~Huan \etal Intergalactic Gamma-Ray Cascade Model}

\authors{H.~Huan$^{1}$, T.~Weisgarber$^{1}$}
\afiliations{$^1$Enrico Fermi Institute, University of Chicago, Chicago, IL 60637, USA}
\email{hhuan@uchicago.edu}

\abstract{Primary gamma rays emitted by extragalactic sources, such as blazars, will generate electromagnetic cascades in intergalactic space. These cascades proceed via electron-positron pair production and inverse Compton scattering on cosmic background radiation, mainly the cosmic microwave background (CMB) and extragalactic background light (EBL) fields. The existence of an extragalactic magnetic field (EGMF) could deflect electron-positron pair trajectories and scatter the cascade photons, possibly creating a halo around the source while suppressing the cascade flux collected by a detector. We develop a semi-analytic model for the cascade process and apply it to combined GeV-TeV data on high-frequency-peaked BL Lacertae objects (HBLs) from the \textit{Fermi} Large Area Telescope (LAT) and ground-based Cherenkov telescopes, comparing observation results with model predictions using a robust statistical framework. Lower limits with different confidence levels on the field strength of the EGMF derived from this procedure are discussed under various assumptions about the source livetime.}
\keywords{ Gamma-Ray Astronomy, Extragalactic Magnetic Field, Electromagnetic Cascade, BL Lacertae Objects, Extragalactic Background Light, \textit{Fermi} Gamma-Ray Space Telescope, VERITAS }

\begin{document}
\maketitle

\section{Introduction}
Extragalactic blazars emit gamma rays in both high-energy (HE, 100 MeV
$\lesssim E\lesssim$ 300 GeV) and very-high-energy (VHE, $E\gtrsim$ 100 GeV)
bands. Due to the existence of the extragalactic background light (EBL),
which spans over the optical to far-infrared wavelength range, gamma rays
with energies above 10 GeV may be absorbed and
produce electron-positron pairs~\cite{Gould1967}.
An electromagnetic cascade then develops via inverse Compton scattering
of the $e^{\pm}$ pairs on the cosmic microwave background (CMB) and subsequent
secondary pair productions. In the presence of an extragalactic magnetic field (EGMF), the charged pairs
in the cascade will be deflected, spreading the cascade photons
in both the spatial and temporal distributions~\cite{Aharonian1994,Plaga1995}. The characteristic angular
spread could create an apparent halo around the point source, and the time delay of cascade photons could appear in the
observation of gamma-ray bursts or flaring blazars. Therefore, gamma-ray
astronomy of extragalactic sources provides a useful probe into the EGMF
strength and configurations.

According to previous studies~\cite{Neronov2007,Elyiv2009} this method will
be sensitive to EGMF below $10^{-14}$ Gauss, much lower than what any other
measurement has achieved. For instance, from Faraday rotation
measurements on extragalactic radio sources~\cite{Kronberg1994,Blasi1999} or analysis of CMB anisotropy~\cite{Barrow1997,Durrer2000}
only an upper limit $\sim 10^{-9}$ Gauss on the field strength is obtained.
The new EGMF window below $10^{-14}$ Gauss is particularly interesting for
the understanding of astrophysical magnetic fields. A
primordial field within the window could be responsible for generating the
galactic and intra-cluster magnetic fields~\cite{Widrow2002}, and its own origin can be related
to either the inflationary era or phase transitions in the early universe~\cite{Grasso2001}. On
the other hand, if the EGMF strength turns out to be zero, the astrophysical
fields would have to be coming from seed fields produced locally via the
Biermann battery mechanism or related processes~\cite{Gnedin2000}. Hence a lower limit on the EGMF instead of a
measurement could be already useful in clarifying the origin of all
the magnetic fields we have in the universe today.

One way for obtaining the
lower limit, using gamma rays as a probe, would be to compare the cascade flux
from VHE emission of a source with the actual HE observed spectrum~\cite{Murase2008,Neronov2010}. If the
HE measured flux is lower than the zero-field cascade flux prediction,
the EGMF would have to be non-zero to dilute the cascade photons into a
spreading angle and thus suppress the collected flux. Deriving an EGMF lower
limit in this way requires a realistic model of the electromagnetic cascade
correlated with EGMF. Existing simplified analytic models ~\cite{Tavecchio2010} and
Monte Carlo simulations~\cite{Dolag2011} set a lower bound for the field strength at
$10^{-16}$ to $10^{-15}$ Gauss assuming the studied sources to be active with
unlimited livetime  or $10^{-19}$ to $10^{-17}$ Gauss for the sources to be
active for only $\sim 3$ years of simultaneous HE and VHE observations~\cite{Tavecchio2011,Dermer2011,Taylor2011}. In this work we model the cascade semi-analytically and
use our model predictions to place a lower limit on the EGMF using a systematic
framework, with the blazar RGB J0710+591 as an example for application and data analysis.

\section{Model Description}
\begin {figure}[!t]
	\begin {center}
		\includegraphics[width=\columnwidth,clip,trim=0 0 0 2.4in]{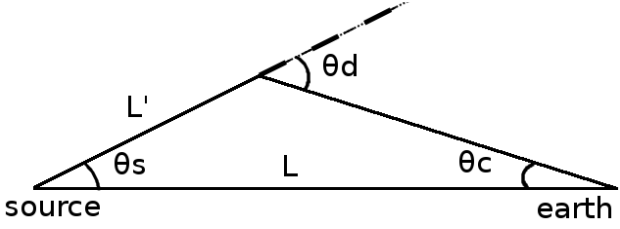}
		\caption{Sketch of the gamma-ray cascade geometry.}
		\label{fig:geometry}
	\end {center}
\end {figure}

The geometry of the cascade is shown in Fig.~\ref{fig:geometry}. Primary
gamma rays emitted by a blazar at distance $L$ from the earth are absorbed
after going through distance $L'$. The electron-positron pairs get deflected
by the EGMF to angle $\theta_d$ and upscatter CMB to secondary photons directed
toward the detector at an
incidence angle $\theta_c$. The emission angle of the primary photon at the
source with respect to the line of sight is $\theta_s=\theta_d-
\theta_c$. The difference in path length between the secondary photon and a
direct photon that goes from the source to the observer in a straight line
is
\begin {equation}
	\Delta L=c\Delta T=L'+\sqrt{L^2+L'^2-2LL'\cos\theta_s}-L\label{eq:delay}
\end {equation}
where $\Delta T$ is the time delay of the secondary photon and $c$ is the speed of light.

For a primary photon with energy $\epsilon$, we assume the electron and
positron produced in the pair-production process each carry $\epsilon/2$.
Then the pair starts to lose energy continuously via inverse Compton scattering
on the CMB while being deflected by the EGMF at the same time. An electron with
Lorentz factor $\gamma_e$ on average scatters CMB photons to energy
$4\gamma_e^2\epsilon_0/3$, where $\epsilon_0$ is the average CMB photon energy
$\sim 6.4\times 10^{-4}$ eV~\cite{Blumenthal1970}. The energy loss process is then described by
\begin {equation}
	\frac{d\gamma_em_ec^2}{dt}=-\frac{4}{3}\epsilon_0n_{\text{CMB}}c\sigma_T\gamma_e^2\label{eq:loss}
\end {equation}
where $n_{\text{CMB}}$ is the CMB photon number density $\sim 411 \text{cm}^{-3}$ and $\sigma_T$ is the Thomson cross section $\sim 6.65\times10^{-25} \text{cm}^{2}$.
The Lorentz deflection process is, on the other hand
\begin {equation}
	\frac{d\theta_d}{dt}=\frac{c}{r_l}=\frac{eB}{\gamma_em_ec}\label{eq:def}
\end {equation}
where $r_l=\gamma_em_ec^2/eB$ is the Larmor radius, $m_e$ is the electron rest
mass, and we have assumed the EGMF $\textbf{\textit{B}}$ to be perpendicular to the
electron momentum. Eqs.~\ref{eq:loss} and \ref{eq:def} combine to give the
deflection angle for an electron/positron to go from Lorentz factor $\gamma_{e0}$
to $\gamma_e$:
\begin {equation}
	\theta_{d0}=\frac{3}{8}\frac{eB}{\epsilon_0n_{\text{CMB}}\sigma_T}
	\left(\gamma_e^{-2}-\gamma_{e0}^{-2}\right)\label{eq:angle}
\end {equation}
which is generalized to
\begin {equation}
\theta_d=\arccos\left(\sin^2\theta_f\cos\theta_{d0}+\cos^2\theta_f\right)\label{eq:angles}
\end {equation}
when the angle $\theta_f$ between $\textbf{\textit{B}}$ and the
electron momentum is other than $\pi/2$.

Combined with the geometry in Fig.~\ref{fig:geometry} Eq.~\ref{eq:angles} uniquely determines $\theta_s$ and $\theta_c$ for a given
set of $B$, $\gamma_{e0}$, $\gamma_e$, $L'$, and $\theta_f$, provided that
$\theta_c<\pi/2$. The number of secondary photons between energies $E$ and
$E+dE$ produced by the electron going from $\gamma_e+d\gamma_e$ to $\gamma_e$
can be calculated from the CMB spectrum, replacing the CMB photon energy
with $3E/(4\gamma_e^2)$:
\begin {eqnarray}
	dN(E,\gamma_e)&=&cdt\sigma_T\frac{27\pi E^2}{8\gamma_e^4}
	\frac{dE}{h^3c^3\gamma_e^2\left(e^{3E/4(\gamma_e^2kT)}-1\right)}\nonumber\\
		&=&\frac{81\pi E^2m_ed\gamma_e}{32h^3c\gamma_e^8\epsilon_0n_{\text{CMB}}}\frac{dE}{e^{3E/(4\gamma_e^2kT)}-1}\label{eq:diff}
\end {eqnarray}
where $h$ is the Planck constant, $k$ the Boltzmann constant, and $T$ the CMB
temperature at 2.73 K. Integrating over $\gamma_e$, $\gamma_{e0}$ (or
equivalently $\epsilon$), $L'$, and averaging over $\theta_f$ gives the
differential secondary photon flux for a certain field strength $B$ as
\begin {eqnarray}
	& &\frac{dN(E)}{dE}=\frac{81\pi E^2m_e}{16h^3c\epsilon_0n_{\text{CMB}}}
	\int \frac{d\gamma_e}{\gamma_e^8\left(e^{3E/(4\gamma_e^2kT)}-1\right)}\nonumber\\
	&\times&\int_0^{\pi/2} d\theta_fg(\theta_f)\int d\epsilon\int dL'\frac{e^{-L'/
	\lambda(\epsilon)}}{\lambda(\epsilon)}f(\epsilon,\theta_s)\nonumber\\
	&\times&\exp\left(-\sqrt{L^2+L'^2-2LL'\cos\theta_s}/\lambda(E)\right)\label{eq:spectrum}
\end {eqnarray}
where $\lambda(\epsilon)$ is the mean free path of a gamma-ray photon at
energy $\epsilon$, depending on the specific EBL profile. $g(\theta_f)$ is
the probability distribution of $\theta_f$, which is $\sin\theta_f$ for a
randomly pointing field. $f(\epsilon,\theta_s)$ is the intrinsic
spectrum of the source, and we integrate over $\epsilon$ starting from
$2\gamma_em_ec^2$ to 200 TeV, as primary photons beyond that energy are mostly
absorbed within 1 Mpc away from the blazar, and quickly deflected away by the
relatively large magnetic field there. The upper limit on $\gamma_e$ is $100$ TeV$/(m_ec^2)$ and a lower limit at $10^5$ is also placed as a practical matter for the
numerical integration, since there is negligible CMB density beyond 3 meV and
we are not interested in secondary flux below 100 MeV.
The integration limits on $L'$ are enforced through observational cuts on $\Delta T$
and $\theta_c$ via geometry shown in Fig.~\ref{fig:geometry}.

The blazar intrinsic emission $f(\epsilon,\theta_s)$, the photon mean
free path $\lambda(\epsilon)$, and the EGMF directional profile $g(\theta_f)$ are the inputs to the cascade model. In practice
we model the blazar emission as boosted isotropic radiation~\cite{Urry1995}
\begin {eqnarray}
	f(\epsilon,\theta_s)&=&f_0(1-\beta\cos\theta_s)^{-\alpha-1}\epsilon^{-
	\alpha}e^{-\epsilon/E_0}\nonumber\\
		&+&f_0(1+\beta\cos\theta_s)^{-\alpha-1}\epsilon^{-\alpha}
		e^{-\epsilon/E_0}\label{eq:intrinsic}
\end {eqnarray}
where the second term models a counter jet. To obtain a conservative prediction
on the cascade flux we choose the optical depth profile of~\cite{Franceschini2008} which
is relatively transparent for VHE gamma rays. $g(\theta_f)$ is taken as $\sin\theta_f$
as we have no prior assumption on the EGMF configuration.

\section {Model Application and EGMF Constraint}
\begin {figure}[!t]
	\begin {center}
		\includegraphics[width=\columnwidth]{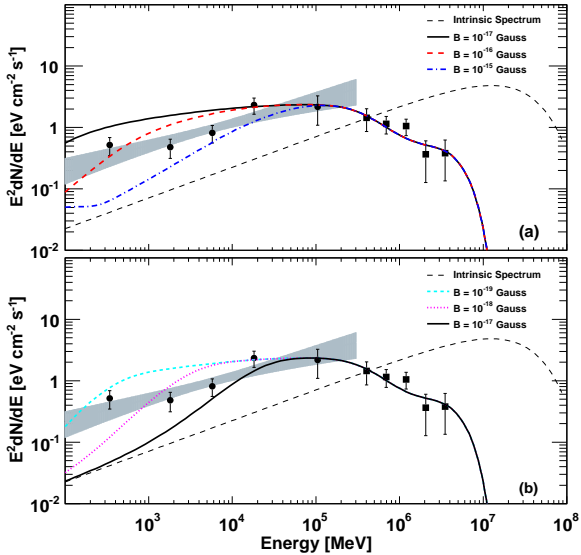}
		\caption{Measured spectral points of RGB J0710+591 and predicted spectra for different assumptions. Squares: VERITAS data. Shaded area: \textit{Fermi} 68\% confidence band. Circles: \textit{Fermi} spectrum. Panel a: Total fluxes for the source with unlimited livetime. Panel b: Total fluxes for the source with 3-year livetime.}
		\label{fig:spectrum}
	\end {center}
\end {figure}

As an example we consider the high-frequency-peaked BL Lacertae object (HBL)
RGB J0710+591 located at redshift $z=0.125$. The predictions of the total
flux as a sum of both the direct and cascade photons within the instrument
point-spread function (PSF) normalized to the observed data are shown in
Fig.~\ref{fig:spectrum} for different EGMF strengths and assumptions
on source livetime, with $\alpha=1.5$, $\Gamma=1/\sqrt{1-\beta^2}=10$, and $E_0=25$ TeV in Eq.~\ref{eq:intrinsic}. The VHE data are from VERITAS measurements~\cite{VERITAS_RGBJ0710} and the HE
data points are extracted from public \textit{Fermi} Large Area Telescope (LAT) data between August
2008 and January 2011 using unbinned likelihood analysis in the
\textit{Fermi} Science Tools v9r18p6 with the instrument response
functions (IRFs) P6\_V3\_DIFFUSE\footnote{Near the completion of this proceeding the \textit{Fermi} team released an updated IRF P6\_V11\_DIFFUSE. We re-analyzed the LAT data with the new IRF and applied the cascade model within the same analysis framework to find that the final constraint on EGMF strength was not significantly affected.}, galactic diffuse emission model gll\_iem\_v02
and isotropic background model isotropic\_iem\_v02\footnote{http://fermi.gsfc.nasa.gov/ssc/}. The $\sim$ 3-year period with
simultaneous HE-VHE data sets a lower limit on the livetime of this source.

\begin {figure}[!t]
	\begin {center}
		\includegraphics[width=\columnwidth]{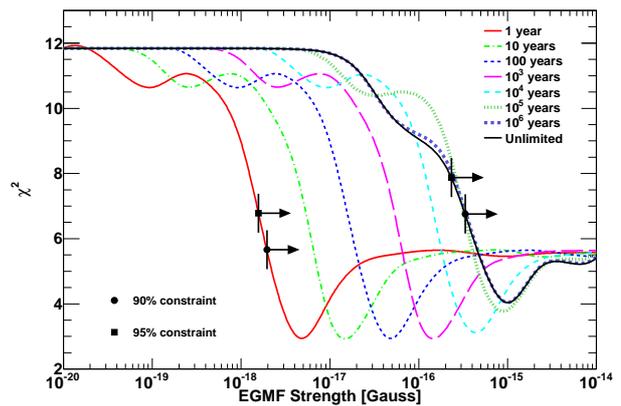}
		\caption{Best-fit $\chi^2$ versus EGMF strength $B$ at different
		livetimes for RGB J0710+591. 90\% and 95\% lower limits on $B$
		are indicated for the 1-year and unlimited livetime cases but
		omitted at other livetimes for clarity.}
		\label{fig:chi2}
	\end {center}
\end {figure}

By requiring that the HE total flux not exceed the \textit{Fermi} LAT measured
spectrum, we can roughly see the EGMF strength $B$ has a lower limit between
$10^{-16}$ and $10^{-15}$ Gauss for the unlimited livetime case, or between
$10^{-18}$ and $10^{-17}$ Gauss for the livetime assumption of 3 years. A more
systematic lower limit could be derived by fitting the total flux to the
measured data points with free parameters of normalization $f_0$, index $\alpha$
and cutoff energy $E_0$. To take into account of complexities in blazar modeling (e.g., \cite{Bottcher2008})
we also fit with a broken power law at 80 GeV and add one more free parameter
$\alpha_{\text{break}}$ below the break energy, only to find that the resulting
lower limit is not greatly affected. Restricting the range of $\alpha$ and
$\alpha_{\text{break}}$ to be no harder than the physically motivated hardness limit 1.5~\cite{Malkov2001,HESS_2006}
and the range of $E_0$ to be (0.1 TeV, 100 TeV],
we plot the minimum $\chi^2$ from the fit as a function of EGMF
strength in Fig.~\ref{fig:chi2} at various source livetime limits. As
expected all the curves converge at low EGMF strengths or large livetimes.

\begin {figure}[!t]
	\begin {center}
		\includegraphics[width=\columnwidth]{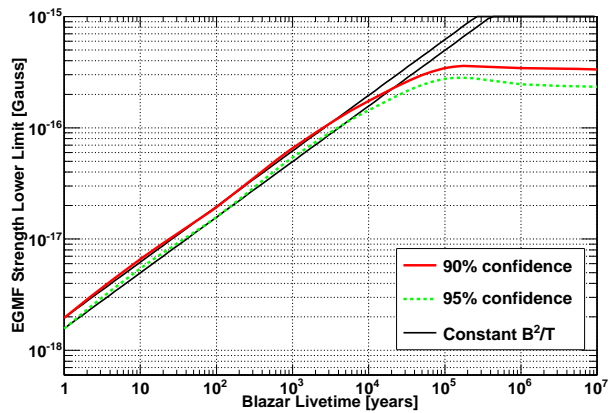}
		\caption{EGMF lower limit as a function of blazar livetime for
		two different confidence levels derived from RGB J0710+591 data.
		Solid black lines: constant $B^2/T$ lines agreeing with the
		curves for small livetimes.}
		\label{fig:limit}
	\end {center}
\end {figure}

The EGMF lower limits at different confidence levels are derived by finding
the point where $\chi^2$ exceeds its minimum value in each curve by $\Delta\chi^2$
in Fig.~\ref{fig:chi2}, which is just a variant of the profile likelihood
method for determining confidence intervals. Two sample confidence levels
(90\% and 95\%) are given by requiring $\Delta\chi^2$ to be 2.72 and 3.84~\cite{James2006},
respectively. We show these lower limits versus the blazar livetime in Fig.~\ref{fig:limit}.
At livetimes below $\sim 10^{-4}$ years, i.e., when the $\Delta T$ constraint
is dominating over the \textit{Fermi} LAT PSF constraint on $\theta_c$,
we have the EGMF lower limit scaling with $\Delta T$ as $B\sim\sqrt{\Delta T}$,
consistent with Eqs.~\ref{eq:delay} and \ref{eq:angle}. The nominal lower limit
at 95\% confidence level is $B\gtrsim 2\times 10^{-16}$ Gauss if the source has
unlimited livetime and $B\gtrsim 3\times 10^{-18}$ Gauss if the source has the
minimum livetime $\sim 3$ years.

\section {Conclusions}
The 95\% lower limits on EGMF strength inferred by the semi-analytic cascade
model for both source livetime assumptions are consistent with the results from full Monte Carlo simulations of
the cascade in~\cite{Taylor2011} on the same blazar, demonstrating that the
model correctly takes into account the important geometrical and physical
aspects involved in the cascade calculation. The limits are conservative as
we choose a comparably transparent EBL model and also assume the electron-positron
pairs to be always going in one coherent magnetic field domain, which is only valid when the EGMF coherence
length $\lambda_B\gtrsim 1$ Mpc. If $\lambda_B\lesssim 1$ Mpc instead, the charged
pairs would random walk through the EGMF domains~\cite{Ichiki2008}, and the deflection angle would
become smaller, causing a larger cascade flux and a more stringent lower limit
on $B$.

Because we ignore cosmological expansion and evolution, the model
is only applicable to sources within a redshift of 0.2. However most of the TeV
blazars already detected are located well in this range \footnote{http://tevcat.uchicago.edu} so we have a pool of
candidate sources with moderate size. We choose RGB J0710+591 as an example
because it has simultaneous data in GeV and TeV and does not have any
variability detected in either energy band. The caveat with this one-source
study is that the EGMF along the line of sight to the source may not be
representative of the overall field strength, e.g., when there is a filament
of intra-cluster magnetic field along the direction. Therefore a statistically
more reliable constraint should be obtained by studying an unbiased set of
blazars and using the systematic framework presented here to combine the results
on single sources. As the gamma-ray telescopes continue to monitor the sky,
more and more blazars with simultaneous GeV-TeV baseline spectra will be
observed and we will be approaching the goal of probing the all-sky EGMF.

{The computations used in this work were performed on the Joint Fermilab - KICP Supercomputing Cluster, supported by grants from Fermilab, Kavli Institute for Cosmological Physics, and the University of Chicago.\\We thank T. Arlen, V. V. Vassiliev, and S. P. Wakely for helpful discussions and comments throughout this work.}

\clearpage

\end{document}